\documentclass[twocolumn,prb,showpacs,preprintnumbers]{revtex4}
\usepackage{graphicx}% Include figure files
\usepackage{dcolumn}% Align table columns on decimal point
\usepackage{bm}% bold math
\usepackage{epsfig}
\begin{document}

%\preprint{APS/123-QED}

\title{A thermodynamic treatment of the glass transition}% Force line breaks with \\

\author{U. Buchenau}
\email{buchenau-juelich@t-online.de}
\affiliation{%
Institut f\"ur Festk\"orperforschung, Forschungszentrum J\"ulich\\
Postfach 1913, D--52425 J\"ulich, Federal Republic of Germany}%
\date{April 17, 2012}% It is always \today, today,

\begin{abstract}
In undercooled liquids, the anharmonicity of the interatomic potentials causes a volume increase of the inherent structures with increasing energy content. In most glass formers, this increase is stronger than the vibrational Gr\"uneisen volume expansion and dominates the thermal expansion of the liquid phase. For a gaussian distribution of inherent states in energy, the generic case, this implies a $1/T^2$-temperature dependence of the additional thermal expansion and the additional heat capacity at zero pressure. The corresponding compressibility contribution has the Prigogine-Defay ratio one. In experiment, one finds a higher Prigogine-Defay ratio, explainable in terms of structural volume changes without any energy change. These should always exist, though their influence becomes weak in close-packing systems, at the crossover to soft and granular matter.
\end{abstract}


\maketitle

When an undercooled liquid freezes into a glass at the glass temperature $T_g$, the thermal expansion usually decreases by a factor of two to four \cite{angell,gupta,nemilov,gundermann}. Obviously, the possibility to jump from one possible structure to another leads to a strong thermal expansion. This possibility gets lost as the system enters the glass phase.

In the glass phase, the thermal volume expansion $\alpha_g$ has a textbook explanation \cite{kittel} in terms of the Gr\"uneisen relation for the vibrations
\begin{equation}\label{gruen}
	\alpha_g=\frac{\Gamma c_{Vg}}{B},
\end{equation}
where $c_{Vg}$ is the heat capacity of the glass at constant volume per unit volume, $B$ is the bulk modulus of the glass and the Gr\"uneisen parameter $\Gamma$ describes the volume dependence $\overline{\omega}\propto 1/V^\Gamma$ of the average vibrational frequency $\overline{\omega}$.

%%%%%%%%%%%%%%%%%%%%% begin figure %%%%%%%%%%%%%%%%%%%%%%%%%%%%%%%%%%%%%
\begin{figure} 
\hspace{-0cm} \vspace{0cm} \epsfig{file=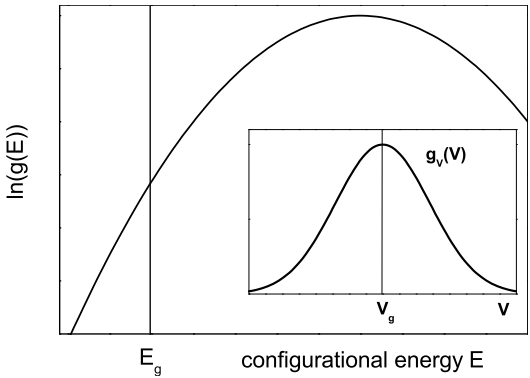,width=7cm,angle=0} \vspace{0cm}\caption{The distribution of inherent states in energy at zero pressure. The inset shows the volume distribution of the states at $E_g$ around the average value $V_g$.}
\end{figure}
%%%%%%%%%%%%%%%%%%%%% end figure %%%%%%%%%%%%%%%%%%%%%%%%%%%%%%%%%%%%%%%

The present paper intends the derivation of a similarly simple-minded relation for the additional thermal volume expansion $\Delta\alpha=\alpha_l-\alpha_g$ of the undercooled liquid. The additional expansion is ascribed to the volume increase of the inherent states with increasing energy content (the inherent states are the structurally stable states which one obtains when cooling the momentary configuration of the liquid to the temperature zero \cite{stillinger}). If the vibrational Gr\"uneisen expansion is the same for all inherent states, one can separate the two influences. As will be seen, this simple concept leads to a physical understanding of the large Prigogine-Defay ratio at the glass transition \cite{nemilov,gundermann}.

In first order, the anharmonicity of the interatomic potential leads to a linear relation between the structural energy $E$ and the inherent structure volume $V$ at constant pressure. Let $E_g$ be the average structural energy and $V_g$ be the average inherent structure volume at the glass temperature $T_g$ and zero pressure. Then one can linearize 
\begin{equation}\label{uv}
	V=V_g+a(E-E_g)
\end{equation}
for not too large deviations of $E$ from the value $E_g$. The anharmonicity factor $a$, an inverse pressure, is a measure for the anharmonicity of the interatomic potential. Note that the compressibility of a single inherent state does not enter in this relation; it is a glass property, which does not require transitions between different inherent states to become measurable. 

To get the partition function $Z$, one has to integrate the density $g(E)$ of the inherent states per atom over the configurational energy $E$ (see Fig. 1). With the linear relation of eq. (\ref{uv}), one then has
\begin{equation}\label{z}
	Z=\int_{-\infty}^\infty g(E)\exp(-\beta E(1+pa))dE,
\end{equation}
where $\beta=1/kT$ and the right factor under the integral, the Boltzmann factor, contains the pressure product $pa$.

One can calculate the average structural energy $\overline{E}$ per atom and the average squared structural energy $\overline{E^2}$ per atom at zero pressure
\begin{equation}\label{ebar}
	\overline{E}=\frac{1}{Z}\int_{-\infty}^\infty Eg(E)\exp(-\beta E)dE
\end{equation}
and
\begin{equation}\label{e2bar}
	\overline{E^2}=\frac{1}{Z}\int_{-\infty}^\infty E^2g(E)\exp(-\beta E)dE.
\end{equation}

With $N$ atoms in the volume $V$, the configurational part $\Delta c_p$ of the heat capacity at zero pressure per unit volume is given by
\begin{equation}\label{dcp}
	\Delta c_p=\frac{N}{V}\frac{\partial\overline{E}}{\partial T}=\frac{N}{VkT^2}(\overline{E^2}-\overline{E}^2),
\end{equation}
and since $V=V_g+a(N\overline{E}-E_g)$, the configurational part $\Delta\alpha$ of the thermal volume expansion is
\begin{equation}
	\Delta\alpha=a\frac{N}{V}\frac{\partial\overline{E}}{\partial T}=a\Delta c_p.
\end{equation}

Finally, the compressibility contribution from the possibility to change the average structural energy is easily calculated from the derivative of the average energy per atom with respect to pressure at constant temperature and the pressure zero
\begin{equation}\label{dk}
	\Delta\kappa_{PD}=a\frac{N}{V}\frac{\partial\overline{E}}{\partial p}=\frac{Na^2}{VkT}(\overline{E^2}-\overline{E}^2).
\end{equation}
The compressibility has the index $PD$, because it satisfies the Prigogine-Defay relation for a second-order phase transition \cite{jackle}
\begin{equation}\label{prigo}
	\frac{\Delta c_p\Delta\kappa_{PD}}{(\Delta\alpha)^2T}=\frac{\overline{\Delta H^2}\ \ \overline{\Delta V^2}}{(\overline{\Delta H\Delta V})^2}=1.
\end{equation}
Here $\Delta H$ and $\Delta V$ are the additional enthalpy and volume fluctuations from the structural energy changes, respectively. For completely correlated enthalpy and volume fluctuations (an implicit assumption of eq. (\ref{uv})), the Prigogine-Defay ratio is one.

According to numerical simulation results \cite{heuer} the density $g(E)$ of the inherent states tends to be a gaussian
in the configurational energy E, with the maximum $E_0$ of the gaussian at a value higher than $kT_g$ (see Fig. 1)
\begin{equation}
	g(E)=g_0\exp\left(\frac{-(E-E_0)^2}{2\overline{w^2}}\right),
\end{equation}
where $\overline{w^2}$ is the mean squared deviation of $E$ from $E_0$. In this simple case
\begin{equation}\label{cpt}
	\Delta c_p=\frac{\overline{w^2}}{Vk_BT^2},
\end{equation}
and the thermal expansion at zero pressure
\begin{equation}\label{alt}
	\Delta\alpha=\frac{a\overline{w^2}}{Vk_BT^2}.
\end{equation}

%%%%%%%%%%%%%%%%%%%%% begin figure %%%%%%%%%%%%%%%%%%%%%%%%%%%%%%%%%%%%%
\begin{figure}
\hspace{-0cm} \vspace{0cm} \epsfig{file=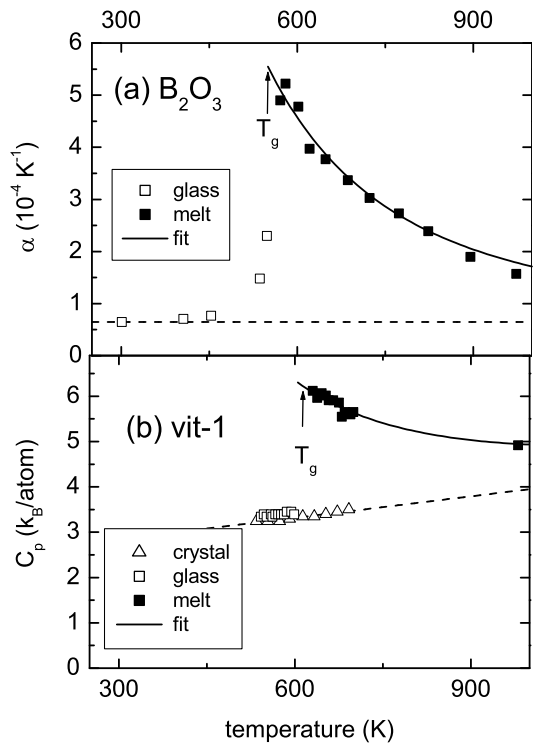,width=7cm,angle=0} \vspace{0cm}\caption{(a) The temperature dependence of the thermal volume expansion \cite{macedo} in B$_2$O$_3$ compared to the sum (continuous line) of a constant Gr\"uneisen term (dashed line) and the structural expansion of eq. (\ref{alt}) (b) The temperature dependence of the heat capacity of the metallic glass vitralloy-1 at ambient pressure \cite{busch} compared to the sum (continuous line) of the vibrational heat capacity of the glass (dashed line) and the inherent structure contribution of eq. (\ref{cpt}).}
\end{figure}
%%%%%%%%%%%%%%%%%%%%% end figure %%%%%%%%%%%%%%%%%%%%%%%%%%%%%%%%%%%%%%%

Though one cannot expect every glass former to have a single gaussian density of inherent states, it turns out to be easy to find substances which have the $1/T^2$-dependence. Fig. 2 shows two examples \cite{macedo,busch}, B$_2$O$_3$ and the metallic glass vitralloy-1. There are more such cases \cite{angell}.

But when one measures the Prigogine-Defay ratio at the glass transition \cite{gupta} in B$_2$O$_3$, one finds a value of 4.7, much larger than unity. This is contrary to the expectation of eq. (\ref{prigo}). Obviously, there must be additional density fluctuations not taken into account by eq. (\ref{dk}).

The inset of Fig. 1 shows the physical reason for these additional density fluctuations: The zero pressure volume $V_g$ at the glass temperature $T_g$ is only the average of the volume of the inherent states at the zero pressure energy $E_g$; one has to reckon with a mean square deviation $\overline{{\rm v}^2}$ of their volume distribution. Therefore, eq. (\ref{uv}) is only valid for the average volume. The volume at constant structural energy has additional fluctuations, which according to the fluctuation-dissipation theorem lead to an additional compressibility $\Delta\kappa_0=\overline{{\rm v}^2}/Vk_BT$. At zero pressure, these additional density fluctuations occur at constant energy and do neither contribute to the heat capacity nor to the thermal expansion. As a consequence, one finds the Prigogine-Defay ratio
\begin{equation}\label{PD1}
	\Pi=\frac{\Delta\kappa_0+\Delta\kappa_{PD}}{\Delta\kappa_{PD}}=\frac{\Delta\kappa}{\Delta\kappa_{PD}},
\end{equation}
where $\Delta\kappa$ is the measured value. $\Pi$ is larger than one if $\Delta\kappa_0$ is larger than zero.

Consider two inherent states with structural energies $E_1$ and $E_2$. At $T_g$ and zero pressure, they have the weighted energy difference $(E_2-E_1)/kT_g$. After application of a small pressure $p$, this increases by the factor $(1+pa)$. To restore the weighted energy difference to its original value, one has to raise the temperature by $\Delta T=paT_g$. Then one returns to the same situation, in particular to the same relaxation time $\tau_\alpha$ of the flow at $T_g$. Thus the pressure dependence of $T_g$ is given by
\begin{equation}\label{eh1}
\frac{\partial T_g}{\partial p}=aT_g=\frac{\Delta\alpha T_g}{\Delta c_p},
\end{equation}
which is in fact the one of the two Ehrenfest relations for second order phase transitions \cite{speedy} found to be valid in glass formers \cite{angell}. The other one differs by the factor $1/\Pi$ from the one for a second order phase transition.

To get an estimate of $\Delta\kappa_0$, consider the structural relaxation processes which bring the shear modulus $G$ of the glass down to zero, the elementary processes of the flow. After applying a small shear strain to an initial inherent state, the shear relaxation flow processes end up in inherent states of the same average structural energy as the initial one. The inherent states at higher energy may be important for the dynamics as intermediate states, but the shear stress release is describable in terms of a sum of final transitions at constant average structural energy. In general, these transitions will not only change the strain state, but the volume as well. Therefore, one has to expect a reduction $\Delta B$ of the bulk modulus $B$ of the glass by
\begin{equation}\label{dbdg1}
	\Delta B=\frac{\delta B}{\delta G}G,
\end{equation}
where $\delta B/\delta G$ is the average coupling ratio of the structural relaxation processes. Thus
\begin{equation}
	\Delta\kappa_0=\frac{1}{B-\Delta B}-\frac{1}{B}
\end{equation}
so one can determine the average coupling ratio $\delta B/\delta G$ from the measurements at $T_g$ via
\begin{equation}\label{dbdg}
	\frac{\delta B}{\delta G}=\frac{B\Delta\kappa(\Pi-1)}{1+B\Delta\kappa(\Pi-1)}\frac{B}{G}.
\end{equation}

The question is: Why do some glass formers like B$_2$O$_3$ have a large Prigogine-Defay ratio (a strong coupling of the structural relaxation processes to an external compression), while others have a Prigogine-Defay ratio close to unity \cite{gundermann}, i.e. structural relaxations which do not couple to an external compression?

Part of the answer to this question has been given in numerical studies \cite{nick,ulfth} of different interatomic potentials. These studies have a relaxation time range of nanoseconds, in the best case microseconds. Therefore they do not discriminate between vibrations and structural relaxation, but calculate the total enthalpy-volume correlation.
They find a strong correlation in the Lennard-Jones potential, a slightly weaker but still strong correlation in the MGCU-potential applicable to metallic glasses, but a rather weak correlation for hydrogen bonded substances. 

Taking the heavily studied \cite{cavagna} Lennard-Jones example, the strong enthalpy-volume correlation (equivalent to a Prigogine-Defay ratio close to unity) at zero pressure is the same as the one for a steep inverse power law potential with an applied external pressure \cite{nick,ulfth}. The inverse power law is $1/r^{18.9}$ ($r$ interatomic distance). This shows that the Lennard-Jones potential is rather close to the hard-sphere case. The applied external pressure needed to hold the atoms together in the equivalent inverse power potential can be estimated from the linear potential term of $2.4\ r$ in Lennard-Jones units at the first coordination shell \cite{nick,ulfth}; it is not small.

The effect of an external pressure on the structural states of the inset of Fig. 1, the ones responsible for the compressibility $\Delta\kappa_0$, is to change their energies. Thus they are no longer at equal energy and begin to contribute to the heat capacity and to the thermal expansion. As a consequence, the Prigogine-Defay ratio diminishes. Its deviation from its zero pressure value should become notable at the critical pressure
\begin{equation}\label{pcrit}
	p_{crit}=\frac{1}{a}=\frac{\Delta c_p}{\Delta\alpha},
\end{equation}
because then the structural states of the inset of Fig. 1 have half the energy-volume coefficient of the other ones.

The strong enthalpy-volume correlation of the Lennard-Jones system extends down to low temperatures in the glass and even in the crystal \cite{nick}, showing that in this case the vibrations reflect the properties of the structural relaxation. This is consistent with the finding $\delta B/\delta G=0$ in a Lennard-Jones glass at zero temperature \cite{wittmer,leonforte}. The instantaneous affine shear deformation modulus $G_\infty$ is a factor of two higher than the final $G$, but $B_\infty=B$. Since $B_\infty$ and $G_\infty$ have the central-force Poisson ratio $\nu=1/3$, this pushes \cite{leonforte} the final $\nu$ up to 0.4. The effect is due to a non-affine motion of the atoms which lowers the shear energy, but does not couple to the compression. The non-affine motion is intimately related to the boson peak and to the tunneling states which dominate the glass behavior at very low temperatures \cite{bggprs,schirmacher} as well as to the plastic modes responsible for the shear thinning in Non-Newtonian flow \cite{procaccia}.

One can define a (numerically accessible) vibrational coupling ratio in the low-temperature glass
\begin{equation}\label{dbdg2}
\frac{\delta B}{\delta G}=\frac{B_\infty-B}{G_\infty-G}.	
\end{equation}
which in the Lennard-Jones case is zero, as well as the one defined in eq. (\ref{dbdg}). There seem to be several examples for an equality even if the ratio is nonzero \cite{bs}. This supports numerical evidence \cite{candelier} for an intimate relation between the soft modes and the structural rearrangements of the undercooled liquid, a property which glass forming liquids seem to share with colloids and granular matter \cite{liu}.

There are some indications that one has a low $\delta B/\delta G$ in metallic glasses as well, though their anharmonic thermal expansion \cite{samwer} $\alpha_lT_g=0.035$ is a factor of ten smaller than the Lennard-Jones one \cite{nick}. Tunneling state measurements \cite{bellessa} show an unusual factor of four weaker coupling of the tunneling states to longitudinal than to transverse waves, consistent with the complete absence of a coupling to the compression. Similarly, one does not see the Ioffe-Regel limit in the x-ray Brillouin scattering from longitudinal waves in a metallic glass \cite{ruocco} at the boson peak, in contrast to measurements in other glass formers \cite{benoit}, but in agreement with a soft-sphere simulation \cite{schober}, which only shows the Ioffe-Regel limit for the transverse waves. The findings indicate that one has $\delta B/\delta G$ close to zero in the rather harmonic metallic glasses as well as in the anharmonic Lennard-Jones case, in agreement with the numerical finding \cite{nick} for the MGCU potential.

There are two possible reasons for this weak compression-relaxation coupling in close packing substances: (i) the attractive part of the potential acts as a critical pressure in both Lennard-Jones systems and metallic glasses (ii) the elementary structural relaxation processes in close packing couple only very weakly to the compression. The second possibility is supported by two atomic models for the structural relaxation in close packing, the interstitial \cite{granato,inter} and the gliding triangle \cite{buscho}, both of which couple only to the shear.

To conclude, the paper presents a thermodynamic description of the undercooled liquid which allows to calculate the additional thermal expansion, the additional heat capacity and the additional compressibility above $T_g$ from the properties of the inherent states. The description provides a physical explanation for the large measured Prigogine-Defay ratios at the glass transition. According to this explanation, a Prigogine-Defay ratio of unity is equivalent to a zero coupling of the structural relaxation processes to an external compression. In glass forming systems, this case is the exception rather than the rule. The zero coupling occurs in close packing at zero pressure. Inspite of intense numerical studies, the reason is not yet fully clear.

Helpful discussions with Michael Ohl and Herbert Schober are gratefully acknowledged.

\end{document}